\documentclass[%
superscriptaddress,
preprint,
 amsmath,amssymb,
 aps,prd 
]{revtex4-2}

\usepackage{graphicx}
\usepackage{dcolumn}
\usepackage{bm}
\usepackage{orcidlink}
\usepackage{amsmath}
\allowdisplaybreaks[1]

\begin{document}

\title{Gravitational wave memory and quantum Michelson interferometer}

\author{Zhong-Kai Guo\orcidlink{0000-0002-6548-0952}}
\email{guozhk@amss.ac.cn}
\affiliation{Beijing Institute of Space Mechanics and Electricity, China Academy of Space Technology, Beijing 100094, China}%
\affiliation{Institute of Applied Mathematics, Academy of Mathematics and Systems Science, Chinese Academy of Sciences, Beijing 100190, China }%

\author{Xiao-Yong Wang}
\affiliation{Beijing Institute of Space Mechanics and Electricity, China Academy of Space Technology, Beijing 100094, China}%


\date{\today}

\begin{abstract}
We examined the output of a quantum Michelson interferometer incorporating the combined effects of nonlinear optomechanical interaction and time-varying gravitational fields. Our findings indicate a deviation from the standard relationship between the phase shift of the interferometer's output and the amplitude of gravitational waves. This deviation, a slight offset in direct proportionality, is associated with the gravitational wave memory effect under the conventional settings of interferometer parameters. Furthermore, the results suggest that consecutive gravitational wave memory, or the stochastic gravitational wave memory background (SGWMB), contributes not only to the classical red noise spectrum but also to a quantum red noise spectrum through this new mechanism. This leads to a novel quantum noise limit for interferometers, which may be crucial for higher precision detection system. Our analysis potentially offers a more accurate description of quantum interferometers responding to gravitational waves and applies to other scenarios involving time-varying gravitational fields. It also provides insights and experimental approaches for exploring how to unify the quantum effects of macroscopic objects and gravitation.

\end{abstract}
\maketitle


\section{Introduction}
Memory effect refers to a physical phenomenon that the spacetime structure cannot revert to its initial state after the passage of gravitational waves \cite{PhysRevD.45.520,frauendiener1992note,Favata_2010,PhysRevD.103.043005}, and is linked to gravitational wave sources. The Minkowski spacetime structure before the arrival of gravitational waves is distinct, or supertranslated, from that after the emission of gravitational waves near null infinity,  causing the change of metrological standard. Specifically, it implies that the relative distance between two test objects cannot return to their initial conditions after being influenced by gravitational waves. This effect has not yet been experimentally confirmed \cite{PhysRevD.101.083026,PhysRevD.104.023004,NANOGrav_2023vfo}. However, it may be detectable in future generations of interferometers \cite{PhysRevLett.120.141102}, space-based gravitational wave observatories \cite{Islo2019qht}, or pulsar timing array projects \cite{madison2014assessing}.

At the classical level, this phenomenon and the corresponding response of a classical Michelson interferometer are well studied. Whereas the expected  enhancement in measurement sensitivity in future ground-based gravitational wave detectors approaches and even surpasses the standard quantum limit (SQL) \cite{Punturo_2010,Hild_2011,Danilishin:2019dxq,Reitze:2019iox}, the test masses and light in an interferometer are all inherently quantum in nature, necessitating a more comprehensive quantum modeling of the interferometer. Furthermore, although the classical Michelson interferometer has achieved significant success in the detection of gravitational waves \cite{PhysRevLett_116,Rowan_2016,PhysRevD.88.043007} and then serves as a versatile tool for probing various theories of gravitation and different frameworks that unify quantum and gravity through astrophysical or cosmological processes \cite{PhysRevD.94.084002,Barcelo_2017,PhysRevLett.121.251103,PhysRevLett.126.041302,PhysRevD.103.044031,PhysRevD.108.064038}, the Michelson interferometer itself, in the quantum level, can also serve as a testing bed guiding quantum gravity \cite{Zych_2012, PhysRevLett.110.170401,Armata2017Quantum,Qvarfort:2017jym,Pang2018Quantum,Galley2022nogotheoremnatureof}. The quantum description of Michelson interferometer includes the appropriate incorporation of gravitational effects in the quantum evolution, particularly the quantum coupling between the light, mirrors and gravitational waves, which may introduce non-negligible correctional effects. Additionally, the interaction between quantum interferometers and gravity can give rise to intricate macroscopic quantum phenomena, such as complex entangled states in macroscopic mirrors \cite{PhysRevLett.88.120401,PhysRevLett.91.130401,Mueller-Ebhardt:2007ehs,Yang2010Sagnac,PhysRevA.81.052307}, thereby enlightening fundamental questions in quantum mechanics. 

The current research on quantum interferometers with optomechanical coupling and time-varying gravitational field, specifically the analysis of radiation-pressure quantum noise and SQL, involves semi-quantized discussions and fully quantum ones. The former 
 only quantize the light while the radiation pressure is treated as a classical random force acting on a classical mirror \cite{PhysRevLett.45.75,PhysRevD.23.1693,PhysRevD.65.022002}. Full quantization, on the other hand, treats both light and the mirror as quantum entities, but requires linear-system operators and applicable to linear measurement theory \cite{RevModPhys.52.341,braginsky1995quantum, Danilishin_2012}. However, in the general optomechanical coupling terms \cite{PhysRevA.47.3173, PhysRevA.49.4055,PhysRevA.51.2537}, such as $\hat{a}^\dag \hat{a} \hat{x}$, the powers of the canonical coordinates and momentum exceed two, resulting in a nonlinear system. The standard approach linearizes it under the phenomenological approximation $\hat{a} \rightarrow \langle a \rangle + \delta\hat{a}$ by retaining only the first-order terms when the mean photon number is extremely high. This method may overlook significant physical effects that are relevant in high-precision measurements and could not be suitable for cases with fewer average photons. Therefore, we rigorously consider the effects of the nonlinear optomechanical coupling and incorporate the quantum effects brought about by time-varying gravitational potentials derived from Bondi-Metzner-Sachs (BMS) theory \cite{bondi1962,sachs1962,penrose1987spinors}.

The aim of the present work is to understand the physical implication of gravitational wave memory on a quantum interferometer. At the same time, it is also hoped that the study of this semiclassical interaction between gravity and quantum system will shed light on our understanding on the interface  between general relativity and quantum mechanics, as a humble, small step to enhance our understanding of quantum gravity. 

The structure of this paper is outlined as follows. Section \ref{sec3}  will 
derive the time-varying Hamiltonian term for the mirrors in the quantum interferometer due to the gravitational tidal force induced by gravitational waves, based on rigorous gravitational wave theory. The gravity-mirror coupling and the optomechanical coupling will be combined in section \ref{sec4} to derive the output results of the quantum Michelson interferometer.  In section \ref{secDisscusion} we discuss an instance of correction to the output of an interferometer and explore the novel quantum noise limits induced by a stochastic gravitational wave memory background. We conclude in section \ref{secConclusions}.

\section{Setting up the Hamiltonian formalism}\label{sec3}
In this section, we present the expression for the geodesic deviation equation for  testing mass in interferometers based on BMS method with the Bondi-Sachs (BS) coordinate $(u,r,\theta,\phi)$ \cite{PhysRevD.95.084048,PhysRevD.103.044012,PhysRevD.103.043005} and at the same time work out the tidal force terms to be inserted into the interaction Hamiltonian. 

Consider the spacetime generated by an isolated gravitational source. As a gravitational wave detector is stationed far away spatially from the source and wait long enough time for the gravitational waves of the source to be detected. In the spacetime picture, the detection of gravitational waves takes place at null infinity. The timelike geodesics of the detectors intercepts the waves at null infinity of the spacetime. 

We assume that the orthogonal arms of the interferometer are tangent to the wavefront of the gravitational wave after $3+1$ decomposition and use Newmann-Penrose (NP) formalism \cite{penrose1987spinors,10.1063/1.1665105} with the tetrad choice $\{l^a,n^a,m^a,\bar{m}^a\}$ that the null complex vectors $m^a, \bar{m}^a$ are defined by $\{m^a=1/\sqrt{2}(x^a_1+ix^a_2),\  \bar{m}^a=1/\sqrt{2}(x^a_1-ix^a_2)\}$ where $ x^a_1$ and $ x^a_2$ denote the unit orthogonal vectors of the two arms, $l^a$ is  tangent to the null hypersurface propagating the Weyl curvature, and the null vector $ n^a $ is defined through relation $\tau^a=1/\sqrt{2}(l^a+n^a)$ where $ \tau^a $ is tangent to the geodesics of the test masses.

We have obtained a null tetrad frame adapted to the detector and the wavefront of gravitational waves. By defining $Z_1^a=Lx^a_1,\ Z_2^a=Lx^a_2$ where $L$ denotes the initial arm length of the interferometer and utilizing the above NP tetrad, we decomposed the Jacobi geodesic deviation equations, deriving  approximate expressions from peeling-off theorem \cite{Newman_1962,PhysRevLett.10.66} in the limit of $r \to \infty$ 
\begin{align}\label{EOMFirst}
    \frac{d^2Z_{1,2}}{dt^2}=\frac{1}{m_{1,2}} F_{1,2}(t)+\mathcal{O}\left(\frac{1}{r^2}\right),
\end{align}
in which 
\begin{equation}
    F_{1,2}(t)\equiv -i^{1,2}\frac{i Lm_{1,2}}{2\sqrt{2}r}\left[\Psi_4^0(t)\pm \bar\Psi^0_4(t)\right],
\end{equation}
where $\Psi_4^0$ is the transverse Weyl scalar at source, $m_{1,2}$ denote the mass of the end mirrors of the northern and eastern arms respectively,
$Z_{1,2}$ represent the non-zero components of $Z_{1,2}^a$ oriented align with $\partial \theta$ and $\partial \phi$ respectively and we have used notation $t\equiv u$ to denote Bondi time.

From the second-order temporal derivative in the equation of motion for a mirror, within a cavity of length L, influenced solely by gravitation, we subsequently derive the effective quantum Hamiltonian incorporating the contributions of light, mirrors, and their optomechanical interaction. Law developed an effective Hamiltonian addressing the interaction between a moving mirror and the electromagnetic field, but this model did not account for gravitational influence \cite{PhysRevA.51.2537}. While the combined effects of static gravitation and optomechanical coupling have been explored \cite{Armata2017Quantum,Qvarfort:2017jym}, analyses incorporating time-varying gravitational forces remain unaddressed. Pang {\it et al.} attempted to construct the composite quantum effects when  quantized gravitational waves are coupled with a moving mirror and the light field \cite{Pang2018Quantum}. The Newton-Schrödinger equations can also introduce the gravitational energy of the mirror into quantum Hamiltonian, whereas this approach is subject to considerable controversy \cite{anastopoulos2014problems}. Given the many unresolved issues surrounding the quantization of gravity, we attempt to construct an effective Hamiltonian for the optomechanical system under the influence of gravity in a semi-classical manner, without considering the quantization of the gravitational field but treating gravity as a classical perturbation in the weak-field non-relativistic
limit. The arrival of gravitational waves will bring about temporal changes in the background gravitational field of the interferometer's mirrors, while ignoring the back-action of the mirror on the background spacetime. Following the approach of Law, we still start from the equations of motion to find the effective quantum Hamiltonian.

The existence of electromagnetic field in the cavity and a specific potential field $V(x)$ surrounding the mirror adds terms $-\frac{1}{m}\partial V(x)/\partial x$ and $\frac{1}{2m}\left(\partial A(x,t)/\partial x\right)^2$ on the right-hand side of the Eq. \eqref{EOMFirst} where $A(x,t)$ is the vector potential of the cavity field. After canonical quantization and a series of approximations, including small-displacement approximation and single-mode approximation, the effective Hamiltonian is derived as shown 
\begin{equation}\label{HamThispaper}
H(t)=\hbar\omega\hat{a}^\dag_{}\hat{a}_{}+\hbar \omega_{m}\hat{b}^\dag_{}\hat{b}_{}-\hbar\omega_{m}\kappa_{}\hat{a}^\dag_{}\hat{a}_{}(\hat{b}^\dag_{}+\hat{b}_{}) -k_{} F_{}(t)(\hat{b}^\dag_{}+\hat{b}_{}),
\end{equation}
exhibiting a time-varying nonlinear quantum system with annihilation operators of photon $a$  and mirror's phonon $b$ and corresponding frequency $\omega$ and $\omega_m$, mirror's mass $m$, and coefficients $k\equiv \sqrt{\hbar/2m\omega_m},\ \kappa\equiv k\omega/(\omega_mL)$. 

Regarding the mirror's displacement, there exists a precise solution within the Heisenberg picture
\begin{align}\label{xH}
    \hat{x}_H(t)=&k e^{-i\omega_m t} \hat{b}+k e^{i\omega_m t}\hat{b}^\dag +\frac{\hbar \omega}{mL\omega_m^2} \hat{a}^\dag \hat{a}(1-\cos\omega_m t)\nonumber\\
    &~~~~~~~~~+\frac{1}{m\omega_m }\int_0^t dt'F(t') \sin[\omega_m(t-t')].
\end{align}
However, the photon output from the interferometer, which is our observable and of primary interest, poses a challenge for the solution of photon variables $\hat{a}_H(t)$. This is due to the nonlinear interaction between photon and mirror, rendering an exact solution unattainable. Consequently, only approximate solutions are feasible, which will be the focus of the following section.

\section{quantum Michelson interferometer in time-varying gravitational field}\label{sec4}
\begin{figure}[htbp]
\includegraphics[width=0.4\textwidth]{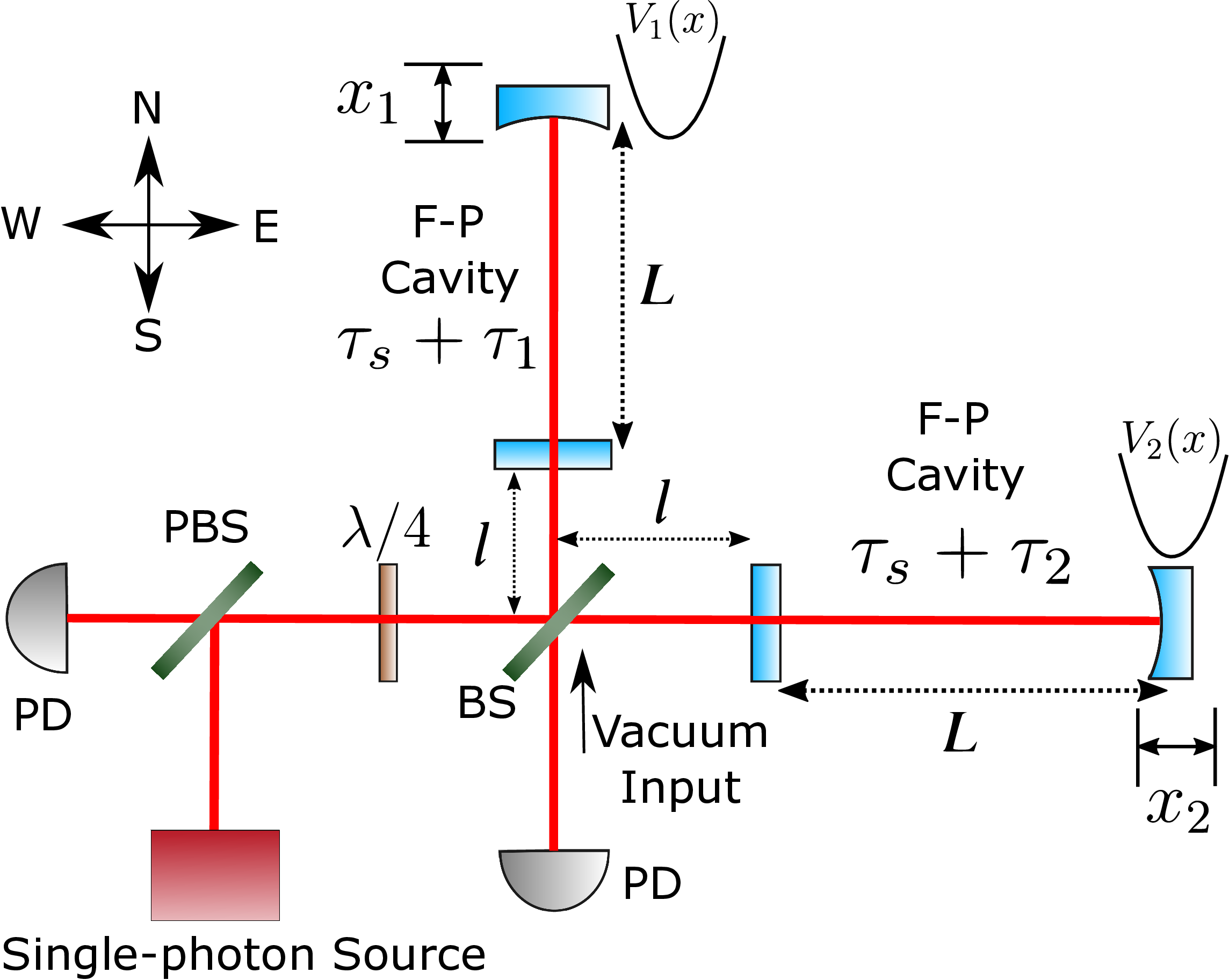}
\caption{\label{fig:Michelson} Quantum Michelson interferometer in time-varying gravitational field.}
\end{figure}
A schematic diagram of the single-photon quantum Michelson interferometer setup is shown in Fig. \ref{fig:Michelson}. Two optical cavities of length $L$ are inserted in the north and east arm of the interferometer respectively, each of which is mounted with a movable end mirror  bounded by a square potential field generated from suspension systems or other electromechanical devices. The mirrors are modelled as harmonic oscillators in small-displacement approximation with natural frequencies $\omega_1,\omega_2$, displacements $x_1,x_2$, mass $m_1,m_2$ and annihilation operators $b_1,b_2$ respectively. We assume $x_{1,2}\ll L$, $ \omega_{1,2}\ll\omega$, where  $\omega$ is the frequency of the photon, leading to the optomechanical coupling Hamiltonians $\hbar\omega x_{1,2} a^\dag_{1,2}a_{1,2}/L$ \cite{PhysRevA.47.3173, PhysRevA.49.4055,PhysRevA.51.2537}, in which $a_{1,2}$ are annihilation operators of the north and the east photon split by the beam splitter respectively. 

For simplicity, we define the notation of optomechanical evolution Hamiltonian $H_{\text{OM1}}\equiv H_{01}-\hbar\omega_1\kappa_{1}a^\dag_{1}a_{1}(b^\dag_{1}+b_{1})$ in which $H_{01}\equiv \hbar\omega a^\dag_{1}a_{1}+\hbar \omega_{1}b^\dag_{1}b_{1}$ is the free evolution term and $\kappa_1\equiv \frac{\omega}{\omega_1 L}\sqrt{\hbar/2m_1\omega_1}$ is a dimensionless optomechanical coupling coefficient. Hence the Hamiltonian describing the north cavity system with moving mirror and time-varying gravitational field is 
\begin{equation}\label{allH}
H_{1}(t)=H_{\text{OM1}}+H_{1t}(t),
\end{equation}
where $H_{1t}(t)\equiv -k_{1}F_1(t)(b^\dag_{1}+b_{1})$ and $k_1\equiv \sqrt{\hbar/2m_1\omega_1}$.
And by substituting subscripts $1\to 2$ immediately gives the  Hamiltonian of the east cavity $H_2(t)=H_{\text{OM2}}+H_{2t}(t)$.

A single photon denoted by number state $|1\rangle_W$ is emitted from a quantum light source and then is split by a beam splitter. In terms of step function
\begin{equation}
u(t_1,t_2)=\begin{cases}1,& t_1<t<t_2,\\
0,&\text{otherwise},
\end{cases}
\end{equation}
 we set the first splitting time is time zero and write the Hamiltonian of north system
\begin{equation}
H_{\text{N}}\!=\!H_{01}u(0,l)\!+\!H_1 u(l,l\!+\!\tau_s\!+\!\tau_1)\!+\!H_{01}u(l\!+\!\tau_s\!+\!\tau_1,2l\!+\!\tau_s\!+\!\tau_1),
\end{equation}
where $l$ is the distance from the splitter to the cavity, $\tau_s=NL/c$ is the photon storage time in the cavity  with number of photon round trips $N$ without gravitational waves, and $\tau_{1,2}\approx -i^{1,2} \cdot iN L/(2\sqrt{2}rc) \left[\sigma^0(t)\pm\bar{\sigma}^0(t)\right] $ are small additions to storage time $\tau_s$ resulting from gravitational waves and $\sigma^0$ relates to the shear of the $(\theta,\phi)$ sphere in BS coordinate \cite{PhysRevD.103.043005}. By similarly substituting the subscripts $1\to2$, we obtain the expression for the east system $H_{\text{E}}$, and the same substitutions can be applied to the following ones.

The evolution operator for $H_{01}$ is trivial $U_{01}(t)=\exp{\left(-i\omega_cta^\dag_1a_1\right)}\exp{\left(-i\omega_1 tb^\dag_1b_1\right)}$ while that for $H_{\text{OM1}}$ is complicated \cite{PhysRevA.56.4175}
\begin{equation}
U_{\text{OM1}}(t)\!=\!e^{-i\omega t a^\dag_1 a_1}e^{iA_1(a^\dag_1 a_1)^2} \!e^{a^\dag_1 a_1\left(z_1b^\dag_1 \!-\!z^*_1b_1\right)}\!e^{-i\omega_{1} tb^\dag_1 b_1},\\
\end{equation}
in which $A_1\!\equiv \!\kappa^2_1\left(\omega_1t\!-\!\sin\omega_1 t\right)$ and $z_1\!\equiv \!\kappa_1 (1\!-\!e^{-i\omega_1 t})$. But $H_1(t)$ is a Hamiltonian with explicit time-dependence, differing from the cases in \cite{Armata2017Quantum,Qvarfort:2017jym}, and therefore the accurate expression of its evolution operator is the time-ordering form $U_{1}(t)=\hat{T}\exp\left(i\int_0^tdt'H_1(t')/\hbar\right)$, which is hardly solved analytically. Fortunately the time-dependent term $H_{1t}(t)$ in $H_1$ can be regarded as perturbation quantity because $F_1(t)\propto 1/r$ with the astronomical distance $r$ from gravitational wave source to the interferometer. So it is reasonable to apply  time-dependent perturbation analysis for the time evolution of quantum states determined by $H_1$. We transform Schrödinger picture into interaction picture, write down the Dyson series solution, and then obtain the time-evolution operator  of $H_1$ at the first order approximation 
\begin{align}
&U_{1}(\tau_s+\tau_1)\equiv U_{1}(l,l+\tau_s+\tau_1)=U_{\text{OM1}}(\tau_s+\tau_1)\nonumber\\
-\frac{i}{\hbar}&U_{\text{OM1}}(l+\tau_s+\tau_1)
    \int^{l+\tau_s+\tau_1}_{l}H^I_{1t}(t')dt'\ U^\dag_{\text{OM1}}(l),
\end{align}
where $H^I_{1t}(t)$ is the perturbation term expressed in interaction picture
\begin{equation}
H^I_{1t}(t)\!\equiv\! -k_1\!F_1\!(t)\!\!\left[e^{i\omega_1t}b^\dag_1\!+\!e^{-i\omega_1t}b_1\!+\!2\kappa_1(1\!-\!\cos\omega_1\! t)a^\dag_1a_1\!\right],
\end{equation}
derived from $U_{\text{OM1}}^\dag(t)H_{1t}(t)U_{\text{OM1}}(t)$ with the help of the Baker-Campell-Hausdorff formula.

We now contemplate a scenario where the mirrors initially reside in their ground states, denoted as $|0\rangle_1|0\rangle_2$, and the south input of beam splitter is in a vacuum state, represented by $|0\rangle_S$, culminating in the initial state of photon-mirror system being $|\psi\rangle=|0\rangle_S|1\rangle_W|0\rangle_1|0\rangle_2$. Based on the quantum model of beam splitter \cite{zeilinger1981general,PhysRevA.40.1371,gerry2005introductory} and by setting its type as $J_2$ and 
balanced (50:50) $\theta=-\pi/2$, the initial quantum state transforms as follows after the first splitting
\begin{equation}\label{intialState}
|\psi\rangle=\frac{1}{\sqrt{2}}(|1\rangle_N|0\rangle_E+|0\rangle_N|1\rangle_E)|0\rangle_1|0\rangle_2.
\end{equation}

From the photon state transformation formula Eq. \eqref{splittingEq} during the second beam splitting 
we derive the total quantum state $|\psi\rangle$ Eq. \eqref{finalState}  after the second beam splitting to the first order approximation of $H_{1t}$ and $H_{2t}$.

Thus the expected photon count on the south side $\langle N_S\rangle\equiv\langle \psi|N_S|\psi\rangle$ or west side $\langle N_W\rangle$ yields the southern output or western output  of the quantum interferometer respectively 
{\small \begin{equation} \label{resultsStrict}\begin{aligned}
&\langle N_{S,W}\rangle=\frac{1}{2}\!\mp\!\frac{1}{2}e^{-\frac{1}{2}\left[|z_1(\tau_s+\tau_1)|^2+|z_2(\tau_s+\tau_2)|^2\right]}\\
   &\times \cos\left[\omega(\tau_1\!-\!\tau_2)\!+\! A_2(\tau_s\!+\!\tau_2)\!-\!A_1(\tau_s\!+\!\tau_1)\right]\\
	&\pm\!\frac{1}{2\hbar}\!e^{-\!\frac{1}{2}\!\left[|z_1(\tau_s\!+\!\tau_1)|^2\!+\!|z_2(\tau_s\!+\!\tau_2)|^2\right]}\big[k_1M_1(\tau_s\!+\!\tau_1)\!-\!k_2M_2(\tau_s\!+\!\tau_2)\big]\\
  &\times \sin\big[\omega(\tau_1-\tau_2)+A_2(\tau_s+\tau_2)
	-A_1(\tau_s+\tau_1)\big],
\end{aligned}
\end{equation}
}where $z_i(t)\!\equiv\!\kappa_i(1\!-\!e^{-i\omega_i t}),\ |z_i(t)|^2\!=\!2\kappa_i^2(1\!-\!\cos\omega_i t),\ A_i(t)\!\equiv\!\kappa_i^2(\omega_i t\!-\!\sin\omega_i t),\ 
	M_i(t)\!\equiv\! 2\text{Re}[C_i(t)z_i(l+t)]\!+\!2D_i(t)=2\kappa_i \int_l^{l+t} F_i(t')\left[1\!-\!\cos\omega_i(l+t-t')\right]dt',\ C_i(t)\!\equiv\!\int^{l+t}_l F_i(t')e^{i\omega_it'}dt',\ 
	D_i(t)\!\equiv\!\kappa_i\int_l^{l+t}F_i(t')(1\!-\!\cos\omega_i t')dt',\  i\!=\!1,2,$ and $|z\rangle,\, \text{Re}[...]$ denote the coherent state with complex number eigenvalue  $z$ and the real value of $[...]$ respectively. 

If we set the two mirrors to be identical,  which means
\begin{equation}
\omega_{1,2}=\omega_m,\ k_{1,2}=k,\ z_{1,2}=z,\  A_{1,2}=A,\  \kappa_{1,2}=\kappa,
\end{equation}
and adjust the potential field $V(x)$ of the mirrors, such as suspended system, along with the arm length $L$ and the number of photon round trips $N$, to ensure that the product of its natural frequency $\omega_m$ and the storage time $\tau_s$ constitutes a small quantity that the condition $\omega_m\tau_s\ll 1$ prevails, then the significant terms $M_1(t),M_2(t) $ in Eqs. \eqref{resultsStrict} can be simplified as
\begin{equation}
\begin{aligned}
M_i(t)&=\kappa_i\omega_i^2\int_{l}^{l+t} F_i(t')(t'-t-l)^2dt'\\
&=2\kappa_i\omega_i^2\int_0^{t}dt'''\int_0^{t'''}dt''\int_l^{l+t''}F_i(t')dt' ,\  i=1,2,
\end{aligned}
\end{equation}
where the second formula can be demonstrated to be identical to the first through the application of integration by parts twice.
We further assume that 
$\tau_1,\, \tau_2,\, F_1(t)$ and $F_2(t)$ are significantly smaller compared to $\tau_s$. Consequently, the strict expressions Eqs.\eqref{resultsStrict} can be reduced to the first-order approximation of $\mathcal{O}((\tau_1+\tau_2)/\tau_s)$ (Appendix \ref{perturbative}), denoted as
\begin{align}
\langle N_{S,W} \! \rangle\! \approx\! \frac{1}{2}\! &\mp\! \frac{1}{2}e^{\! -2\kappa^2\left[1-\cos\omega_m\tau_s+\frac{\omega_m}{2}(\tau_1+\tau_2)\sin\omega_m\tau_s\right]}\notag\\
&\times \cos \left\{(\tau_1-\tau_2)\left[\omega-\omega_m\kappa^2(1-\cos\omega_m\tau_s)\right]\right\}\nonumber\\
	&\pm\frac{\omega}{2mL}e^{-2\kappa^2\left[1-\cos\omega_m\tau_s+\frac{\omega_m}{2}(\tau_1+\tau_2)\sin\omega_m\tau_s\right]}\notag\\
&\times \sin\left\{(\tau_1-\tau_2)\left[\omega-\omega_m\kappa^2(1-\cos\omega_m\tau_s)\right]\right\}\notag\\
&\times\!  \int_0^{\tau_s}\!\!\!\!dt'''\!\!\!\int_0^{t'''}\!\!\!\!dt''\!\!\!\int_l^{l+t''}\!\!\!\!\left[F_1(t')-F_2(t')\right]dt',\label{NSFinal}
\end{align}
which are the primary findings of this paper. And the findings express the measurement outputs from photon detectors in a quantum interferometer conducted under the condition $\omega_m \tau_s \ll 1$ and subjected to the influences of a weak time-varying gravitational field, exemplified by gravitational waves. The first two terms represent quantum corrections arising from optomechanical coupling, in contrast to the standard interferometer outputs $\langle N_{S,W}\rangle=1/2\mp 1/2 \cos[\omega (\tau_1-\tau_2)]$. The third term signifies quantum corrections due to  composite effects of nonlinear optomechanical coupling and gravity-mirror coupling. Compared to the results from the previous papers with linear systems, an additional perturbative corrective multiple integral term of the difference in Weyl tensors has emerged and the differential displacement term $\tau_1-\tau_2$ it multiplies  is phase-shifted by exactly $\pi/2$ ($\cos[...]\to\sin[...]$) relative to the standard term, making this result quite novel.

Most importantly, the multiple integral term, as defined, represents the difference of gravitational wave memory effects between the two arms if the duration of the gravitational radiation is less than $\tau_s$
\begin{align}\label{Memory}
&\int_0^{\tau_s}dt'''\int_0^{t'''}dt''\int_l^{l+t''}\left[F_1(t')-F_2(t')\right]dt'\!\notag\\
\!&\approx \!m\!\!\!\int_l^{l+\tau_s}\!\!\!\!\!\!\!\!\!\!\!\!\! \left[\! Z_1(t)\!-\!Z_2(t)\!\right]\!dt\!-\!m\tau_s\!\left[\!Z_1(l)\!-\!Z_2(l)\!\right]\!-\!\frac{m}{2}\tau_s^2\!\!\left[\!\dot{Z}_1\!(l)\!-\!\dot{Z}_2\!(l)\!\right]\notag\\
\!&\approx \!m L(z_1^m-z_2^m)(\tau_s-\tau_e),
\end{align}
where \(\tau_e\) denotes the time \(t = l + \tau_e\) at which the memory effect has progressed to half its magnitude and $z_1^m\equiv \frac{1}{L}Z_1(t)\big|_{l}^{l+\tau_s},\, z_2^m\equiv \frac{1}{L}Z_2(t)\big|_{l}^{l+\tau_s}$ are memory effects induced by this gravitational wave event. In the second approximate equation, we neglect the integrals of the oscillatory terms of the gravitational waves, which are nearly zero, and retain only the integral of the direct current term produced by the memory effect. Additionally, we disregard the values from the initial moments of the gravitational wave event, including the initial arm length difference $Z_1(l)-Z_2(l)$ and the rate of change in arm length $\dot{Z}_1(l)-\dot{Z}_2(l)$ at the time $t=l$. This indicates that the photon detection results from the interferometer will have an added perturbation correction term related to  memory effect, superimposed on the standard results, leading to one of the main discoveries of this research.

Moreover, within the context of typical experimental parameters for gravitational wave interferometers, the condition $\kappa=\sqrt{\hbar\omega^2/2m\omega_m^3L^2} \ll 1 $ generally holds true, thereby allowing further simplification of the output results Eqs. \eqref{NSFinal} to
{
\begin{align}
&\langle N_{S,W} \rangle\!\approx\!\frac{1}{2}\mp\frac{1}{2}\!\cos \omega(\tau_1\!\!-\!\!\tau_2)\pm\frac{\omega}{2}z^{dm} \tau^{dm} \sin\omega(\tau_1\!\!-\!\tau_2),\label{abbrOutput1}
\end{align}}
where $z^{dm}\!\equiv\! z_1^m\!-\!z_2^m$ represents the difference of memory between the northern and eastern arms and $\tau^{dm}\equiv \tau_s-\tau_e$ represents the time difference between the moment when the memory effect  reaches half its magnitude and the instant when the photon exits the interferometer. It implies that on top of the standard response relationship between gravitational wave amplitude difference and interferometric light intensity, there will be an added, $\pi/2$ out-of-phase, minor correction term. The magnitude of the correction is associated with the memory effects $z^{dm}$ of the gravitational wave over a duration $\tau_s$, the integration time for the memory effect $ \tau^{dm}$, the frequency of the photon $\omega$, and the differential displacement $\tau_1-\tau_2$.

\section{discussion} \label{secDisscusion}
\subsection{Modification of the Proportional Relationship Between Output Phase and Strain Difference}
To elucidate the extent to which this correction impacts the standard relation $\langle N_S\rangle\!=\!1/2\!-\!1/2\cos\omega(\tau_1\!-\!\tau_2)$, we substitute the following parameters that the frequency of  photon $\omega=282 \, \text{THz} \cdot 2\pi$, the arm length  $L=4 \, \text{km}$, the number of photon round trips  $N=150$, the amplitude strength of the gravitational waves $|\Delta L / L| \approx 10^{-21}$, and the frequency of the mirrors' phonon  $\omega_m = 1 \, \text{Hz} \cdot 2\pi$, thereby fulfilling the conditions $\omega_m\tau_s\approx 0.01\ll 1$ and $\kappa\approx 3.2\times 10^{-8}\ll 1$. We also assume that the difference of memory is $ z^{dm}\approx10^{-22}$, and the integration time for the memory is $\tau^{dm}\approx \frac{1}{2}\tau_s$. With these parameters, the expected photon count on the south side has the estimated numerical expression given by
\begin{equation}\label{numerEquation}
\langle N_S \rangle\approx\frac{1}{2}-\frac{1}{2}\cos \omega(\tau_1-\tau_2)+8.9\times 10^{-11}\cdot\sin \omega(\tau_1-\tau_2).
\end{equation}
Although the coefficient $8.9\times 10^{-11}$ of the correction term appears exceedingly small, it has a significant, non-negligible impact because for small values of $\tau_1-\tau_2$, $\sin\omega(\tau_1-\tau_2)$ approximates to the first power of $\omega(\tau_1-\tau_2)$, and $1-\cos\omega(\tau_1-\tau_2)$ approximates to the second power of $\omega(\tau_1-\tau_2)$, thereby rendering the whole correction term consequential.

\begin{figure}[htbp]
\includegraphics[width=0.5\textwidth]{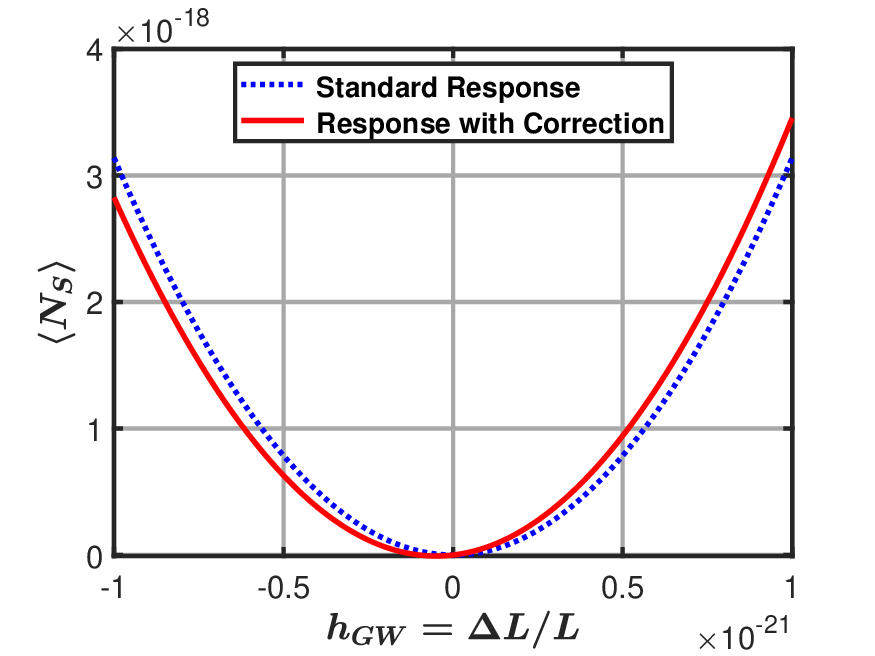}
\caption{\label{fig:standardvsMemory} Correction to the standard response in a quantum Michelson interferometer arising from the combined effects of nonlinear optomechanical coupling and time-varying gravity-mirror coupling, in which $h_{GW}\approx (\tau_1-\tau_2)c/(NL)$ represents the typical strain of gravitational waves.}
\end{figure}

The response curve of the south-side photon detector is shown in Fig.~\ref{fig:standardvsMemory}. The blue dotted curve represents the standard response curve without considering the correction, while the red solid line represents the response curve with the correction. The x-axis represents different strength of  gravitational waves, and the y-axis represents the average photon count on the south-side detector for single-photon input. 
The standard data processing procedure for gravitational wave detection involves converting the average photon count to phase, and then converting the phase to gravitational waves' strain. This is reasonable in the absence of a memory correction term. However, as we can see in the figure, the correction induced by the memory effect causes a slight shift to the standard cosine response curve. The greater the magnitude of the gravitational waves, the larger this deviation. 
\begin{figure}[htbp]
\includegraphics[width=0.5\textwidth]{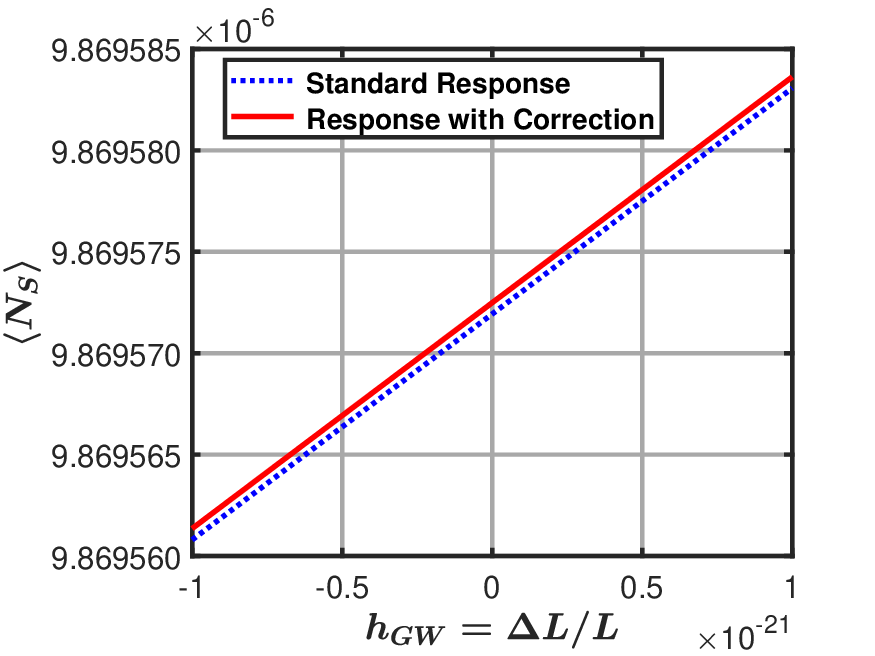}
\caption{\label{fig:standardvsMemorOffset} The response curves of the interferometer influenced by a preset phase offset $\varphi_0 \approx 2\pi \cdot 10^{-3}\, \text{rad}$. The quadratic relationship is converted to a linear one, thereby significantly enhancing sensitivity. The magnitude of the correction no longer increases with the amplitude of the gravitational waves. Instead, it approximates a constant shift.}
\end{figure}

Due to the weak nature of the quadratic response curve, which makes gravitational waves difficult to detect, experimental homodyne readout schemes typically implement an offset to the interferometer \cite{Aasi_2015}. This adjustment converts the response curve from a quadratic to a linear dependency, significantly enhancing sensitivity. Numerically, this is equivalent to adding a fixed initial phase, $\varphi_0$, to the phase component of the output result's trigonometric functions. Specifically, in Eq. \eqref{numerEquation}  $\omega(\tau_1-\tau_2)$  should be replaced with $\omega(\tau_1-\tau_2) + \varphi_0$. To illustrate the impact of this offset, we configured the interferometer with a fixed initial phase of $\varphi_0 \approx 2\pi \cdot 10^{-3}\, \text{rad}$ and redrew the response curve, as shown in Fig. \ref{fig:standardvsMemorOffset}. It depicts the response curves of the interferometer influenced by a preset phase offset, where the blue dotted line represents the standard response curve, and the red solid line illustrates the response curve with quantum gravity correction effects considered. It is clear from the graph that setting a specific offset in the interferometer indeed transforms the response curve from a parabolic-like to a linear-like curve, optimizing the quadratic dependence into a linear one, thereby enhancing the response capability. However, deviations from the standard response due to the correction remain. Importantly, the magnitude of these corrections no longer increases with the strength of the gravitational wave. And instead, there is a small, constant shift across the curve. Mathematically, this constant shift arises when the phase shift induced by the offset, $\varphi_0$, is much greater than the phase shift caused by the gravitational waves, satisfying the condition $\omega(\tau_1-\tau_2)\ll \varphi_0 \ll1$. Under such circumstances, the expression $\sin[\omega(\tau_1-\tau_2) + \varphi_0] \approx \sin \varphi_0+\mathcal{O}([\omega(\tau_1-\tau_2)])$ holds. Therefore, when presetting a certain offset in the interferometer, the output will still exhibit this correction effect.

For the more realistic response curves in quantum scenario, if we still follow the standard procedure of converting the average photon count to phase and then to gravitational waves' strain, it would actually result in a slight misjudgment of the gravitational waves' strain. When the duration of the gravitational wave event exceeds \(\tau_s\) or the conditions  $\omega_m\tau_s\ll1$ and $\kappa\ll 1$ for approximation are not satisfied, it is necessary to apply the more rigorous results given in Eqs. \eqref{resultsStrict} instead of the aforementioned approximate expression.

\subsection{Quantum Noise Limit Induced by Consecutive Memory}
To illustrate the quantum disturbance inflicted on the mirrors by the measured photons and the ensuing issues related to the SQL, we postulate the detection of such photon at the southern detector, resulting in the collapse of state Eq. \eqref{finalState} into an entangled state of mirrors 
{\small
\begin{align}
&\frac{1}{2}\!e^{\!\!-\!i\omega(2l\!+\!\tau_s\!+\!\tau_1)}\!e^{\!iA_1}\!|z_1e^{\!\!\!-i\omega_1 \!l\!}\rangle_1\!|0\!\rangle_2\!\!-\!\!\frac{1}{2}e^{\!\!-i\omega(2l\!+\!\tau_s\!+\!\tau_2)}e^{\!iA_2}\!|0\rangle_1|\!z_2e^{\!\!-i\omega_2 l}\!\rangle_2\notag\\
 &-\frac{i}{\hbar}\frac{1}{2}\Bigg\{e^{-i\omega(2l+\tau_s+\tau_1)}e^{iA_1}\notag\\
 &\bigg\{k_1C_1(\tau_s+\tau_1)e^{-i\omega_1(2l+\tau_s+\tau_1)}b^\dag_1|z_1e^{-i\omega_1 l}\rangle_1|0\rangle_2\notag\\
&+\!k_2C_2(\tau_s\!+\!\tau_1)e^{\!-\!i\omega_2\!(2l\!+\!\tau_s\!+\!\tau_1)}\!\!|z_1e^{\!-\!i\omega_1\!l}\rangle_1\!|1\rangle_2\notag\\
&+k_1\big[C^*_1(\tau_s\!+\!\tau_1)e^{i\omega_1(l+\tau_s+\tau_1)}z_1\!+\!M_1\big]|z_1e^{-i\omega_1 l}\rangle_1|0\rangle_2\bigg\}\notag\\
	&+\!\!e^{\!\!-i\omega(2l\!+\!\tau_s\!+\!\tau_2)}e^{\!iA_2} \!\bigg\{k_2C_2(\tau_s\!\!+\!\!\tau_2)e^{\!\!-i\omega_2(2l\!+\!\tau_s\!+\!\tau_2)}|0\rangle_1b^\dag_2|z_2e^{\!\!-i\omega_1 l}\rangle_2\notag\\
	&+\!k_1\!C_1\!(\tau_s\!+\!\tau_2)e^{\!-i\omega_1\!(2l\!+\!\tau_s\!+\!\tau_2)}\!|1\rangle_1|z_2e^{\!-i\omega_2l}\rangle_2\notag\\
&+\!\!k_2\big[C^*_2(\tau_s\!+\!\tau_2)e^{i\omega_2(l+\tau_s+\tau_2)}z_2\!+\!M_2\big]|0\rangle_1|z_2e^{-i\omega_2 l}\rangle_2\!\!\bigg\}\!\!\Bigg\},	
\end{align}
which will subsequently evolves according to the Hamiltonians 
$\hbar\omega_i b_i^\dag b_i+k_iF_i(t)(b_i^\dag+b_i),\, i=1,2$, until the entrance of the next photon into the optical cavity and then evolves in accordance with Eq. \eqref{allH}. At this point, however, the initial state of mirrors is no longer the ground state as Eq. \eqref{intialState}, but rather the evolved entangled state. This transformation causes the subsequent average output of photons of the interferometer to be influenced by the previous photons, a phenomenon represents a quantum back-action effect, diverging from the description of treating the coupling of light and mirrors as linearized radiation pressure fluctuation exerted on the mirrors and solving the equation of motion for the mirrors \cite{PhysRevD.65.022002,Danilishin_2012}.

Generally, we posit that the initial quantum state of the light field and the mirror is $|\alpha\rangle_{light}|\beta\rangle_m$ where $\alpha$ and $\beta$ denote the eigenvalues of corresponding coherent states of light and mirror, respectively. Then the average displacement of the mirror, as induced by the combined effects of light and a singular gravitational wave memory following Eq. \eqref{xH}, is
\begin{align}
\langle x\rangle(\!{\tau_s}\!)\!=&2k|\beta|\cos\left[\omega_m \tau_s- \text{arg}(\beta)\right]\!+\frac{\hbar\omega|\alpha|^2}{mL\omega_m^2}(\!1-\cos\omega_m \!\tau_s)\nonumber\\
&+\frac{1}{m\omega_m}\!\!\int_0^{\!\tau_s} \!\!\!\!\!dt'F(t') \sin[\omega_m(\tau_s-t')],
\end{align}
in which $\text{arg}(\beta)$ refers to the argument of the complex number $\beta$. 
We assume that the natural frequency of the mirror is sufficiently low fulfilling $\omega_m \tau_s\ll 1$, thereby closely resembling that of a free mirror, allowing us to derive a free-mass approximate expression 
\begin{equation}
 \langle x\rangle({\tau_s})\!=\!2k|\beta|\!\cos\left[ \text{arg}(\beta)\right]\!+\frac{\hbar\omega |\alpha|^2\tau_s^2}{2mL}+\frac{1}{m}\int_0^{\tau_s}\!\!dt'\int_0^{t'}\!\!F(t'')dt'',   
\end{equation}
in which the first term indicates the expected initial position of the mirror, and the second term represents the displacement of the mirror resulting from the momentum transfer of expected $|\alpha|^2$ photons colliding N times, while the third term corresponds to the displacement of the mirror attributed to the displacement memory effect of gravitational waves, which are analogous to the classical scenario. 

We then focus on the impacts attributable to gravitational wave memory.
After $n$ gravitational wave memory events, the average displacement and the corresponding power spectrum density (PSD) $S(f)$ of the mirror become
\begin{equation}
    \langle x\rangle =\sum_{j=1}^n Z^m_j, \quad S^{dm}(f)= \frac{L^2Q}{f^2},
\end{equation}
where $Z^m_j\equiv \frac{1}{m}\int_0^{\tau_s}dt'\int_l^{l+t'}F^{(j)}(t'')dt''$ denotes the displacement memory effect resulting from the j-th gravitational wave event, the inverse square relationship of the power spectrum originates from the classical Brownian motion induced by multiple instances of random displacement memory effects, and $Q$ is a coefficient that is dependent upon the astrophysical and cosmological context. This corresponds to the background noise induced by classical stochastic gravitational wave memory background (SGWMB) \cite{Zhao:2021zlr}. 

However, for photons' variables  $a_N$ and $a_S$, due to the nonlinear interaction between the photon and mirror, no exact solution is available. Consequently, the proportional relationship between mirror displacement and the output phase of the interferometer inevitably fails to hold. And it is impossible to directly deduce the expression for the output phase of the interfered photons from mirrors' displacement above. Instead, starting from the mean output of photon numbers from a single gravitational wave event Eq. \eqref{abbrOutput1}, we phenomenologically express the average number of photon outputs following $n$ instances of stochastic gravitational wave memory events as
\begin{equation}
\langle N_S\rangle=\frac{1}{2}-\frac{1}{2}\cos\left(\delta\phi^{(n)}\right)+\frac{\omega}{2 }\sin\left(\delta\phi^{(n)}\right) \sum_{j=1}^n  z^{dm}_j\tau_j^{dm},  
\end{equation}
in which $\delta\phi^{(n)}\equiv \frac{\omega NL}{c}
\sum_{j=1}^nz^{dm}_j$ denotes the cumulative phase change in the output light induced by the total memory. The first two terms are the output of the traditional stochastic memory effect and the third  term  describes a new type of fundamental noise, specifically the measurement noise formed by the interaction of the stochastic gravitational wave memory background with the quantum objects of the interferometer, constituting a fourth type of quantum noise originated from the combined effects of quantum mechanics and gravity, independent of the Heisenberg limit of the mirrors, photon shot noise, and photon radiation pressure noise, thus establishing a novel quantum noise limit. This can be termed as the Memory Quantum Noise (MQN). In accordance with the conventional linear back-projection method $\delta\langle N_S\rangle\sim \delta\phi\sim\delta  x$ \cite{Danilishin_2012}, which correlates mean photon number of the output with displacement, then the displacement noise attributable to the memory effect can be equivalently described as the additional noise of $\langle N_s\rangle$ with PSD
\begin{equation}
    S^{\text{MQN}}(f)\approx \frac{\omega^4 N^2L^2\tau_s^2}{16c^2}\frac{Q^2}{f^2},
\end{equation}
assuming that each $\tau_j^{dm}$ is a random variable uniformly distributed across the interval [0, \(\tau_s\)]. The above equation indicates that the interference caused by memory quantum noise remains a red noise spectrum. It is pertinent to note that these conclusions are preliminary. Due to the complexity involved in accurately computing the effects of a non-linear, time-dependent Hamiltonian on photons under consecutive memory effects, our discussion is qualitative rather than based on rigorous computation. And also as our discussions are predicated on quantum systems, this necessitates an extremely low temperature for the mirrors, thereby imposing stringent requirements on the experimental setup. 

\section{conclusions}\label{secConclusions}
Our primary contribution is the calculation of the detector output for a quantum Michelson interferometer in the presence of both nonlinear optomechanical coupling and time-varying gravity-mirror coupling. Due to the great distance between the interferometer and the gravitational source, we can treat the gravity-mirror coupling as a perturbation and obtain the southern and western outputs of the quantum interferometer given by Eqs. \eqref{resultsStrict}. Selecting interferometer parameters such that \(\omega_m\tau_s \ll 1\) and incorporating the approximation \(\tau_1, \tau_2 \ll \tau_s\), we obtain the approximate results as presented in Eqs. \eqref{NSFinal}, which include a perturbative correction term compared to the optomechanical coupling results. Notably, this correction term is related to the memory effect described by Eq. \eqref{Memory}. 

To clarify the extent to which our results modify the standard detector output, we take the condition $\kappa=\sqrt{\hbar\omega^2/2m\omega_m^3L^2}\ll 1$ that are typically satisfied for interferometer parameters  to further simplify the results to Eqs. \eqref{abbrOutput1}. Upon substituting a specific set of parameters, it becomes evident that a notable, albeit modest, deviation from the standard relation occurs. This indicates that the data processing procedure of inferring instantaneous phase from instantaneous light intensity and then inferring the instantaneous amplitude of the gravitational waves is inherently flawed at the theoretical quantum level. This is a relatively novel example of how the gravitational wave memory effect influences the output of gravitational wave detectors. Besides, when the duration of each gravitational wave is less than storage time $\tau_s$ and there is a continuous influence of a large number of gravitational wave memory effects on the interferometer, we derive a novel type of quantum noise arising from the composite effect of quantum mechanics and gravity. The PSD of this noise remains red. Our findings could potentially provide a reference of significance for the advancement of future high-precision gravitational wave detectors and the field of quantum gravity.

Experimentally, the aforementioned effects become significant only when the mirrors' quantum properties are pronounced. The mirrors must be initially in their ground states at the frequency $\omega_m$ and other noise sources such as gravity gradient noise, seismic noise, and suspension thermal noise should not cause greater disturbances at this frequency. Advancements in the preparation of macroscopic oscillators in quantum ground states \cite{PhysRevLett.99.073601,Abbott_2009,Teufel_2011,Chan:2011ivv} and the the latest advancements in single-photon sources \cite{Esmann_2024,yang_2024} render the empirical examination of gravity-influenced fully quantum interferometers feasible in the near future. Relocating interferometers to space could also significantly reduce gravity gradient noise and seismic noise, leaving thermal noise as the primary concern  \cite{amaroseoane2017laser}. Certainly, for actual ground-based gravitational wave detection, which utilizes Schnupp asymmetry and homodyne or heterodyne readout schemes \cite{Aasi_2015}, the response of light power to gravitational waves is much more complex and requires further analysis. Moreover, investigating the entangled state of macroscopic mirrors and its implications for foundational inquiries in quantum mechanics, as well as considering the incident light in a coherent state, will be pivotal in future research endeavors aimed at yielding more pragmatically significant results. Additionally, our efforts should be directed towards developing a comprehensive depiction of a fully quantum interferometer exhibiting   general covariance under gravitational influence \cite{Pang2018Quantum}, with the objective of examining the proper integration of quantum mechanics with gravitation.

\begin{acknowledgments}
We are grateful to Yanbei Chen for sparking the initial motivation and to Yun-Kau Lau for suggesting this research problem to us and subsequent many discussions. We also wish to acknowledge the guidance of Haixing Miao and Zhoujian Cao. This work was supported by the NSFC (Grant No. 11571342).
\end{acknowledgments}

\appendix
\section{}
For the states $|1\rangle_N|0\rangle_E|0\rangle_1|0\rangle_2$ and $|0\rangle_N|1\rangle_E|0\rangle_1|0\rangle_2$, the transformed states prior to the second beam splitting are as follows respectively,
{\small\begin{equation}
\begin{aligned}
&U_{\text{01}}\!(l)U_{1}\!(\tau_s\!+\!\tau_1)U_{\text{01}}\!(l)|1\rangle_N|0\rangle_1U_{\text{02}}\!(l)U_{\text{2}}(\tau_s\!+\!\tau_2)U_{\text{02}}(l)|0\rangle_E|0\rangle_2,\\
&U_{\text{01}}\!(l)U_{\text{1}}\!(\tau_s\!+\!\tau_1)U_{\text{01}}\!(l)|0\rangle_N|0\rangle_1U_{02}\!(l)U_{\text{2}}(\tau_s\!+\!\tau_2)U_{02}(l)|1\rangle_E|0\rangle_2.
\end{aligned}
\end{equation}
}
From the photon state transformation formula during the second beam splitting
\begin{equation}\label{splittingEq}
\begin{aligned}
|1\rangle_N|0\rangle_E\!\stackrel{\text{BS}}{\rightarrow}\!\frac{1}{\sqrt{2}}\!\big(|0\rangle_S|1\rangle_W\!+\!|1\rangle_S|0\rangle_W\!\big),\\
|0\rangle_N|1\rangle_E\!\stackrel{\text{BS}}{\rightarrow}\!\frac{1}{\sqrt{2}}\!\big(\!|0\rangle_S|1\rangle_W\!-|1\rangle_S|0\rangle_W\!\big),
\end{aligned}
\end{equation}
we derive the total quantum state after the second beam splitting to the first order approximation of $H_{1t}$ and $H_{2t}$
{\small
\begin{align}\label{finalState}
&|\psi\rangle=\frac{1}{2}\!e^{\!\!-\!i\omega(2l\!+\!\tau_s\!+\!\tau_1)}\!e^{\!iA_1}\!\big(|0\rangle_S|1\rangle_W\!\!+\!|1\rangle_S|0\rangle_W\!\!\big)\!|z_1e^{\!\!\!-i\omega_1 \!l\!}\rangle_1\!|0\rangle_2\!\notag\\
	&+\frac{1}{2}e^{-i\omega(2l\!+\!\tau_s\!+\!\tau_2)}e^{iA_2}\big(|0\rangle_S|1\rangle_W\!\!-\!|1\rangle_S|0\rangle_W\big)|0\rangle_1|z_2e^{-i\omega_2 l}\rangle_2\notag\\
 &-\frac{i}{\hbar}\frac{1}{2}\Bigg\{e^{-i\omega(2l+\tau_s+\tau_1)}e^{iA_1}\bigg\{k_1C_1(\tau_s+\tau_1)e^{-i\omega_1(2l+\tau_s+\tau_1)}\notag\\
&~~~~~~~~~~~~~~~~~~~~~~~~~\times \big(|0\rangle_S|1\rangle_W+|1\rangle_S|0\rangle_W\big)b^\dag_1|z_1e^{-i\omega_1 l}\rangle_1|0\rangle_2\notag\\
&+\!k_2C_2(\tau_s\!+\!\tau_1)e^{\!-\!i\omega_2\!(2l\!+\!\tau_s\!+\!\tau_1)}\!\big(|0\rangle_S|1\rangle_W\!\!+\!|1\rangle_S|0\rangle_W\!\big)\!|z_1e^{\!-\!i\omega_1\!l}\rangle_1\!|1\rangle_2\notag\\
&+k_1\big[C^*_1(\tau_s\!+\!\tau_1)e^{i\omega_1(l+\tau_s+\tau_1)}z_1\!+\!M_1\big]\big(|0\rangle_S|1\rangle_W\!+\!|1\rangle_S|0\rangle_W\big)\notag\\
 &~~~~~~~~~~~~~~~~~~~~~~~~~~~~~~~~~~~\times|z_1e^{-i\omega_1 l}\rangle_1|0\rangle_2\bigg\}\notag\\
	&-e^{-i\omega(2l+\tau_s+\tau_2)}e^{iA_2} \bigg\{k_2C_2(\tau_s+\tau_2)e^{-i\omega_2(2l+\tau_s+\tau_2)}\notag\\
 &~~~~~~~~~~~~~~~~~~~~~~~~~\times\big(|0\rangle_S|1\rangle_W-|1\rangle_S|0\rangle_W\big)|0\rangle_1b^\dag_2|z_2e^{-i\omega_2 l}\rangle_2\notag\\
	&+\!k_1\!C_1\!(\tau_s\!+\!\tau_2)e^{\!-i\omega_1\!(2l\!+\!\tau_s\!+\!\tau_2)}\!\big(|0\rangle_S|1\rangle_W\!\!-\!\!|1\rangle_S|0\rangle_W\!\!\big)|1\rangle_1|z_2e^{\!-i\omega_2l}\rangle_2\notag\\
&+k_2\big[C^*_2(\tau_s\!+\!\tau_2)e^{i\omega_2(l+\tau_s+\tau_2)}z_2\!+\!M_2\big]\big(|0\rangle_S|1\rangle_W\!-\!|1\rangle_S|0\rangle_W\!\big)\notag\\
&~~~~~~~~~~~~~~~~~~~~~~~~~~~~~~~\times|0\rangle_1|z_2e^{-i\omega_2 l}\rangle_2\bigg\}\Bigg\},	
\end{align}
where $A_{1,2}\equiv A_{1,2}(\tau_s+\tau_{1,2}),\  z_{1,2}\equiv z_{1,2}(\tau_s+\tau_{1,2}),\ M_{1,2}\equiv M_{1,2}(\tau_s+\tau_{1,2})$.
}
\section{Perturbation calculations}\label{perturbative}
Generally, for any real constant $A $ and a small quantity $ \varepsilon $, we reduce the following trigonometric functions to the first order of $ \varepsilon$
\begin{align}
\cos A(1+\varepsilon) &\approx \cos A  - A\varepsilon\sin A, \\
\sin A(1+\varepsilon) &\approx \sin A  + A\varepsilon\cos A.
\end{align}
For the expressions $ \cos\omega_m(\tau_s+\tau_i) = \cos\omega_m\tau_s(1+\tau_i/\tau_s),\ i=1,2 $, since \( \tau_i/\tau_s \) is a small quantity, retaining up to the first order gives
\begin{equation}
\cos\omega_m(\tau_s+\tau_i) \approx \cos\omega_m\tau_s - \omega_m\tau_i\sin\omega_m\tau_s,\ i=1,2.
\end{equation}
Similarly, we have
\begin{equation}
\sin\omega_m(\tau_s+\tau_i) \approx \sin\omega_m\tau_s + \omega_m\tau_i\cos\omega_m\tau_s,\ i=1,2.
\end{equation}
Thus we have the first-order perturbation expansion results related to \( (\tau_1+\tau_2)/\tau_s \)
{\small
\begin{align}
&\!2\!\!-\!\!\cos\!\omega_m\!(\tau_s\!\!+\!\tau_1\!)\!\!-\!\cos\!\omega_m\!(\tau_s\!\!+\!\tau_2\!)\! \!\approx\! 2\!\!-\!\!2\!\cos\!\omega_m\!\tau_s\!\! +\! \omega_m\!(\tau_1\!\!+\!\tau_2\!)\!\sin\!\omega_m\!\tau_s, \label{pertur1} \\
&\!\omega_m\!(\!\tau_2\!\!-\!\!\tau_1\!)\!\!+\!\!\sin\!\omega_m\!(\!\tau_s\!\!+\!\!\tau_1\!)\!\!-\!\!\sin\!\omega_m\!(\!\tau_s\!\!+\!\!\tau_2\!)\! \!\approx\!\! (\!\tau_1\!\!-\!\!\tau_2\!)\!\!\left[\!\omega\!\!-\!\!\omega_m\!\kappa^2\!(\!1\!\!-\!\!\cos\!\omega_m\!\tau_s\!)\!\right]\!. \label{pertur2}
\end{align}
}
Additionally, we have the equation
\begin{align}
&\int_0^{\tau_s+\tau_1}\!\!\!\!dt'''\int_0^{t'''}\!\!\!\!dt''\int_l^{l+t''}\!\!\!\!\!\!\!\!\! F_1(t')dt'\!-\int_0^{\tau_s+\tau_2}\!\!\!\!dt'''\int_0^{t'''}\!\!\!\!dt''\int_l^{l+t''}\!\! \!\!\!\!\!\!\!\!\!F_2(t')dt' \!\notag\\
&=\int_0^{\tau_s}dt'''\int_0^{t'''}dt''\int_l^{l+t''}\!\!\!\!\left[F_1(t')\!-\!F_2(t')\right]dt' \notag \\
&+\int_{\tau_s}^{\tau_s+\tau_1}\!\!\!\!\!\!\!\!dt'''\!\!\int_0^{t'''}\!\!\!\!dt''\!\!\int_l^{l+t''}\!\!\!\!\!\!F_1(t')dt'\!-\!\int_{\tau_s}^{\tau_s+\tau_2}\!\!\!\!\!\!\!\!dt'''\!\!\int_0^{t'''}\!\!\!\!dt''\!\!\int_l^{l+t''}\!\!\!\!F_2(t')dt'\notag\\
&\approx\int_0^{\tau_s}dt'''\int_0^{t'''}dt''\int_l^{l+t''}\!\!\!\!\left[F_1(t')\!-\!F_2(t')\right]dt'.
\end{align}
Since $ F_1, F_2, \tau_1, \tau_2 $ are all small quantities relative to \( \tau_s \), the last two terms are second-order small quantities and can be neglected.
\bibliography{Michelson_quantum_PRD}
\end{document}